\documentclass[prd,aps,showpacs,nofootinbib]{revtex4}%
\usepackage{graphicx}
\usepackage{amsmath}
\usepackage{amsfonts}
\usepackage{amssymb}%
\newcommand{\be}{\begin{equation}}
\newcommand{\ee}{\end{equation}}
\newcommand{\bq}{\begin{eqnarray}}
\newcommand{\eq}{\end{eqnarray}}
\begin{document}
\title{Influence of Lorentz- and CPT-violating terms on the Dirac equation}
\author{Manoel M. Ferreira Jr and Fernando M.O. Moucherek}
\affiliation{Universidade Federal do Maranh\~{a}o (UFMA), Departamento de F\'{\i}sica,
Campus Universit\'{a}rio do Bacanga, S\~{a}o Luiz - MA, 65085-580 - Brasil}

\begin{abstract}
The influence of \ Lorentz- and CPT-violating terms (in "vector" and "axial
vector" couplings) on the Dirac equation is explicitly analyzed: plane wave
solutions, dispersion relations and eigenenergies are explicitly obtained. The
non-relativistic limit is worked out and the Lorentz-violating Hamiltonian
identified in both cases, in full agreement with the results already
established in the literature. Finally, the physical implications of this
Hamiltonian on the spectrum of hydrogen are evaluated both in the absence and
presence of a magnetic external field. It is observed that the fixed
background, when considered in a vector coupling, yields no qualitative
modification in the hydrogen spectrum, whereas it does provide an effective
Zeeman-like splitting of the spectral lines whenever coupled in the axial
vector form. It is also argued that the presence of an external fixed field
does not imply new modifications on the spectrum.

\end{abstract}
\email{manojr@ufma.br, fmoucherek@yahoo.com.br}

\pacs{11.30.Er, 11.10.Kk, 12.20.Fv}
\maketitle

\section{Introduction}

Lorentz covariance, as is well-known, is a good symmetry of the fundamental
interactions comprised in the traditional framework of a local Quantum Field
Theory, from which the Standard Model is derived. \ However, since the
beginning 90%
\'{}%
s, \ Lorentz-violating theories have been proposed as a possible candidate of
signature of a more fundamental physics defined in a higher scale of energy,
not accessible to the present experiments. A pioneering work due to
Carroll-Field-Jackiw \cite{Jackiw} has proposed a CTP-odd Chern-Simons-like
correction term ($\epsilon^{\mu\nu\kappa\lambda}v_{\mu}A_{\nu}F_{\kappa
\lambda})$ to the conventional Maxwell Electrodynamics, that preserves gauge
invariance despite breaking Lorentz and parity symmetries. Some time later,
Colladay \& Kostelecky \cite{Colladay1}, \cite{Colladay2} adopted a quantum
field theoretical framework to address the issue of CPT- and Lorentz-breakdown
as a spontaneous violation \cite{Kostelecky1}-\cite{Kostelecky3}. In this
sense, they constructed the extended Standard Model (SME), an extension to the
Standard Model which maintains unaffected the $SU(3)\times SU(2)\times U(1)$
gauge structure of the usual theory and incorporates the CPT-violation as an
active feature of the effective low-energy broken action. In the broken phase,
the resulting effective action exhibits breakdown of CPT and Lorentz
symmetries at the particle frame, but conservation of covariance under the
perspective of the observer inertial frame. The parameters representing
Lorentz violation are obtained as the vacuum expectation values of some tensor
operators belonging to the underlying theory. The SME incorporates all the
tensor terms that yield scalars (by contracting standard model operators with
Lorentz breaking parameters) in the observer frame.

Timely, it is worthwhile to point out the existence of alternative mechanisms
that bring about equivalent Lorentz-breaking effects. Indeed, noncommutative
field theories \cite{NONC1}-\cite{NONC5} also generate Lorentz-violating terms
of equal structure, able to imply similar effects to the ones of the SME
phenomenology. Another mechanism is varying fundamental couplings
\cite{Vcoupling1}-\cite{Vcoupling3} which amounts to the incorporation of
Lorentz-violating terms in the action as well. In fact, varying couplings
leads to the breaking of temporal and spatial translations, which may be seen
as a particular case of Lorentz breakdown. In a cosmological environment, this
issue may be used to investigate candidate fundamental theories containing a
scalar field with a spacetime-varying expectation value\textbf{, }once the
associated Lorentz-breaking effects may be taken as a signature for an
underlying theory.\textbf{ }Further, Lorentz violation still appears in other
theoretical contexts, involving the consideration of loop gravity
\cite{Loop}-\cite{Loop2} and spacetime foam \cite{Foam}-\cite{Foam2}.

\ The gauge sector of the SME model has been extensively studied in several
works both in (1+3) and (1+2)-dimensions \cite{Coleman}-\cite{Belich5}, with
many interesting results. Concerning the fermion sector, in the context of the
SME, Colladay \& Kostelecky \cite{Colladay1}, \cite{Colladay2} have devised
Lorentz-violating terms compatible with U(1) gauge symmetry and
renormalizability. These terms are explicitly written as below:%
\begin{equation}
\mathcal{L}=-v_{\mu}\overline{\psi}\gamma^{\mu}\psi-b_{\mu}\overline{\psi
}\gamma_{5}\gamma^{\mu}\psi-\frac{1}{2}H_{\mu\nu}\overline{\psi}\sigma^{\mu
\nu}\psi+\frac{i}{2}c_{\mu\nu}\overline{\psi}\gamma^{\mu}\overleftrightarrow
{D}^{\nu}\psi+\frac{i}{2}d_{\mu\nu}\overline{\psi}\gamma_{5}\gamma^{\mu
}\overleftrightarrow{D}^{\nu}\psi, \label{LV1}%
\end{equation}
where the Lorentz-breaking coefficients $v_{\mu},b_{\mu},H_{\mu\nu},c_{\mu\nu
},d_{\mu\nu}$ arise as vacuum expectation values of tensor quantities defined
in an underlying theory. The first two terms are CPT-odd and the others are
CPT-even. Firstly, the fermion sector of the SME model has been investigated
in a general way (by discussing dispersion relations, plane-wave solutions,
and energy eigenvalues). Later, it has been addressed\textbf{ } in connection
with CTP-violating probing experiments , which involve comparative studies of
cyclotron frequencies of trapped-atoms \cite{Cyclotron}-\cite{Cyclotron2},
clock-comparison tests \cite{Clock}, spectroscopic comparison of hydrogen and
antihydrogen \cite{Hydrog}, analysis of muon anomalous magnetic moment
\cite{Muon}, study of macroscopic samples of spin-polarized solids
\cite{Spin}, and so on.

The interest of the present work lays only on the two CPT-odd terms, linked to
the fermion field by an assigned "vector" and "axial vector" coupling,
respectively. The main objective is to examine the effects of the
Lorentz-violating background on the Dirac equation and solutions, focusing on
its nonrelativistic regime and possible implications on the hydrogen spectrum.
Some results concerning this study were already discussed in the literature.
Indeed, the nonrelativistic Hamiltonian associated with Lagrangian (\ref{LV1})
was already evaluated by means of a Foldy-Wouthuysen expansion in refs.
\cite{Hamilton}-\cite{Hamilton2}. Moreover, the corresponding shifts of atomic
levels were perturbatively carried out in a broad perspective in ref.
\cite{Hydrog},\cite{Shifts}. In the present paper, however, the analysis of
the hydrogen spectrum in the presence CPT-odd terms is done in a different way
(more specific, direct and simpler), also including the action of an external
constant magnetic field. The starting point is the Dirac Lagrangian
supplemented by Lorentz and CPT-violating terms. The dispersion relations,
plane-wave solutions and eingenenergies are carried out for each one of the
considered couplings. \ In the sequel, the investigation of the
nonrelativistic limit is performed. This is a point of interest due to its
connection with real systems of Condensed Matter Physics, a true environment
where the presence of a background may be naturally tested. The effect of the
background on the spectrum of hydrogen atom is then evaluated, initially for
the case of the vector coupling, for which it is reported no correction on the
hydrogen spectrum. In the case of the axial vector coupling, the spinor
solutions come out to be cumbersome and the nonrelativistic limit altered. The
Pauli equation is supplemented by terms that effectively modify the spectrum
of the hydrogen in a similar way as the usual Zeeman effect. This sort of
theoretical modification may be combined with fine spectral analysis to set up
precise bounds on the magnitude of the corresponding Lorentz-violating
coefficient. It is still shown that the presence of an external fixed magnetic
field does not lead to new Lorentz-violating effects.

This paper is outlined as follows.\ In Sec. II, it is considered the presence
of the term $v_{\mu}\overline{\psi}\gamma^{\mu}\psi$ in the Dirac Lagrangian.
The modified Dirac equation, dispersion relations, plane-wave solutions and
energy eingenvalues are evaluated. \ The nonrelativistic limit is analyzed and
the corresponding Hamiltonian worked out. In a first order evaluation, it is
shown that the Lorentz-violating terms do not modify the hydrogen spectrum. In
Sec. III the presence of the axial vector term, $b_{\mu}\overline{\psi}%
\gamma_{5}\gamma^{\mu}\psi,$ in the Dirac sector is considered. Again, the
modified Dirac equation, dispersion relations, plane-wave solutions and
eingenvalues are carried out. Finally, the low-energy limit is studied and the
Hamiltonian evaluated. A first order computation shows that the
Lorentz-violating terms contribute to the spectrum hydrogen, causing a Zeeman
splitting of the spectral lines. \ In Sec. IV, one presents the Conclusion and
final remarks.

\section{Lorentz-violating Dirac Lagrangian ("vector" coupling)}

\qquad The most natural and simple way to couple a fixed background \ $\left[
v^{\mu}=(v_{0},\overrightarrow{v})\right]  $ to a spinor field is defining a
vector coupling, given as follows:
\begin{equation}
\mathcal{L}%
\acute{}%
\mathcal{=L}_{Dirac}-v_{\mu}\overline{\psi}\gamma^{\mu}\psi, \label{L1}%
\end{equation}
where $\mathcal{L}_{Dirac}$ is the usual Dirac Lagrangian ($\mathcal{L}%
_{Dirac}=\frac{1}{2}i\overline{\psi}\gamma^{\mu}\overleftrightarrow{\partial
}_{\mu}\psi-m_{e}\overline{\psi}\psi)$ and $v_{\mu}$ is one of the CPT-odd
parameters that here represents the fixed background responsible for the
violation of Lorentz symmetry in the frame of particles. In true, the term
$v_{\mu}\overline{\psi}\gamma^{\mu}\psi$ behaves as a scalar just in the
observer frame, in which $v_{\mu}$ is seen as a genuine 4-vector. The
Euler-Lagrange equation applied on this Lagrangian provides the modified Dirac equation:%

\begin{equation}
\left(  i\gamma^{\mu}\partial_{\mu}-v_{\mu}\gamma^{\mu}-m_{e}\right)  \psi=0,
\label{DiracS1}%
\end{equation}
which corresponds to the usual Dirac equation supplemented by the
Lorentz-violating term associated with the background. The initial task is to
investigate the plane-wave solutions, which may be attained by writing the
spinor in terms of a plane-wave decomposition, $\psi=Ne^{-ix\cdot p}w\left(
p_{\mu}\right)  ,$ where N is the normalization constant and $w\left(  p_{\mu
}\right)  $ is the $\left(  4\times1\right)  $ spinor written in the momenta
space. Taking it into account, eq. (\ref{DiracS1}) is rewritten in momentum
space:
\begin{equation}
\left(  \gamma^{\mu}p_{\mu}-v_{\mu}\gamma^{\mu}-m_{e}\right)  w(p)=0.
\label{DiracS2}%
\end{equation}
It is possible to show that each component of the spinor $w$ satisfies a
changed Klein-Gordon equation which represents the dispersion relation of this
model. In fact, multiplying this equation on the left by $\left(  \gamma^{\mu
}p_{\mu}-v_{\mu}\gamma^{\mu}+m_{e}\right)  ,$ it results:
\begin{equation}
\left(  p\cdot p-2p\cdot v+v\cdot v-m_{e}^{2}\right)  w(p)=0, \label{DR1}%
\end{equation}
whose energy solutions are: $E_{\pm}=v_{0}\pm\sqrt{m_{e}^{2}+(\overrightarrow
{p}-\overrightarrow{v})^{2}}$. Here, one has two different energy values, one
positive $\left(  E_{+}\right)  ,$ another negative $\left(  E_{-}\right)  $.
The negative solution should be reinterpreted as positive-energy
anti-particles. Even after the reinterpretation, the eigenenergies remain
different. This is an evidence of charge conjugation breakdown, as it will be
properly discussed ahead.

Now, the spinors $w(p)$ compatible with such equation should be achieved.
Adopt an explicit representation for the Dirac matrices\footnote{Here, one
adopts the Dirac representation for $\gamma-$matrices: $\gamma^{0}=\left(
\begin{array}
[c]{cc}%
I & 0\\
0 & -I
\end{array}
\right)  ,\ \ \ \gamma^{i}=\left(
\begin{array}
[c]{cc}%
0 & \sigma^{i}\\
-\sigma^{i} & 0
\end{array}
\right)  ,$ $\gamma^{5}=i\gamma^{0}\gamma^{1}\gamma^{2}\gamma^{3}=\left(
\begin{array}
[c]{cc}%
0 & I\\
I & 0
\end{array}
\right)  ,$ with $\sigma^{i}=(\sigma_{x},\sigma_{y},\sigma_{z})$ being the
well-known Pauli matrices.} and writing $w(p)$ in terms of two $2\times1$
spinors ($w_{A}$ and $w_{B})$, the following spinor equations are obtained:%

\begin{align}
w_{A}  &  =\frac{1}{\left(  E-v_{0}-m_{e}\right)  }\overrightarrow{\sigma
}\cdot\left(  {}\overrightarrow{p}-\overrightarrow{v}\right)  w_{B}%
,\label{WA}\\
w_{B}  &  =\frac{1}{\left(  E-v_{0}+m_{e}\right)  }\overrightarrow{\sigma
}\cdot\left(  \overrightarrow{p}-\overrightarrow{v}\right)  w_{A}. \label{WB}%
\end{align}
In order to attain a simple solution, a usual procedure for construction of
plane-wave spinors is followed: a starting form, $\left(
\begin{array}
[c]{c}%
1\\
0
\end{array}
\right)  $ or $\left(
\begin{array}
[c]{c}%
0\\
1
\end{array}
\right)  ,$ for one of them is proposed, so that the other is
straightforwardly derived by means of eqs. (\ref{WA}), (\ref{WB}). These two
$2\times1$ spinors must then be grouped in a single normalized $\left(
4\times1\right)  $ spinor. Following this procedure, after
reinterpretation\footnote{It should be just remembered that the
reinterpretation procedure consists in turning a negative-energy solution into
a positive-energy anti-particle (for which the energy and momentum must be
reverted: $E\rightarrow-E,\overrightarrow{p}\rightarrow-\overrightarrow{p})$%
.}, four independent $\left(  4\times1\right)  $ spinors, $u_{i}$ (particle
solutions) and $v_{i}$ (anti-particle solutions), are attained:%

\begin{equation}
u_{1}(p)=N\left(
\begin{array}
[c]{c}%
1\\
0\\
\frac{\left(  p_{z}-v_{z}\right)  }{E+m_{e}-v_{0}}\\
\frac{\left(  p_{x}-v_{x}\right)  +i\left(  p_{y}-v_{y}\right)  }%
{E+m_{e}-v_{0}}%
\end{array}
\right)  \ ,\ \ \ \ u_{2}(p)=N\left(
\begin{array}
[c]{c}%
0\\
1\\
\frac{\left(  p_{x}-v_{x}\right)  +i\left(  p_{y}-v_{y}\right)  }%
{E+m_{e}-v_{0}}\\
\frac{-\left(  p_{z}-v_{z}\right)  }{E+m_{e}-v_{0}}%
\end{array}
\right)  \ , \label{S1}%
\end{equation}

\begin{equation}
v_{1}(p)=N\left(
\begin{array}
[c]{c}%
\frac{\left(  p_{z}+v_{z}\right)  }{E+m_{e}+v_{0}}\\
\frac{\left(  p_{x}+v_{x}\right)  +i\left(  p_{y}+v_{y}\right)  }%
{E+m_{e}+v_{0}}\\
1\\
0
\end{array}
\right)  \ ,\ \ \ \ v_{2}(p)=N\left(
\begin{array}
[c]{c}%
\frac{\left(  p_{x}+v_{x}\right)  +i\left(  p_{y}+v_{y}\right)  }%
{E+m_{e}+v_{0}}\\
-\frac{\left(  p_{z}+v_{z}\right)  }{E+m_{e}+v_{0}}\\
0\\
1
\end{array}
\right)  , \label{S2}%
\end{equation}
where $N$ is the normalization constant. In the solutions (\ref{S1}),
(\ref{S2}), one of the effects of the background is manifest: to shift the
energy and momentum by a constant: $E\rightarrow E-v_{0},\overrightarrow
{p}\rightarrow\left(  \overrightarrow{p}-\overrightarrow{v}\right)  .$ It is
also instructive to exhibit the energy eigenvalues associated with the four
solutions above. In this case, one can write two eigenvalue equations:
$Hu_{i}=E_{i}^{(u)}u_{i},Hv_{i}=E_{i}^{(v)}v_{i},$ with $i=1,2,$ and
$E_{i}^{(u)}=v_{0}+\left[  m_{e}^{2}+(\overrightarrow{p}-\overrightarrow
{v})^{2}\right]  ^{1/2},E_{i}^{(v)}=\left[  m_{e}^{2}+(\overrightarrow
{p}+\overrightarrow{v})^{2}\right]  ^{1/2}-v_{0}.$ Here, $E_{i}^{(u)}$ stands
for the particle energy whereas $E_{i}^{(v)}$ represents the anti-particle
energy. In the reinterpretation procedure, it was obviously assumed that the
magnitude background is minute near the electron mass ($v_{0}<<m_{e})$,
regarded as a correction effect. This must be so once many experiments
demonstrate the validity of Lorentz covariance with high precision. It should
still be pointed out that these energy values are in agreement with the
similar ones obtained in refs. \cite{Colladay1}-\cite{Colladay2},
\cite{Hamilton}-\cite{Hamilton2}. The attainment of different energies for
particle and anti-particle $\left(  E_{i}^{(v)}\neq E_{i}^{(u)}\right)  $ is
an evidence that the charge conjugation $\left(  C\right)  $ symmetry has been
broken. Indeed, the term $v_{\mu}\overline{\psi}\gamma^{\mu}\psi$ is C-odd and
PT-even, that is, it implies breakdown of charge conjugation, and conservation
of combined PT operation. An ease way to demonstrate such a violation is to
apply the charge conjugation operator $C=i\gamma^{0}\gamma^{2}$ on the
modified Dirac equation, as given in eq. (\ref{DiracS3}). This procedure will
lead to the corresponding Dirac equation for the\ charge conjugate spinor
$\left(  \Psi_{c}=C\Psi^{\ast}\right)  $ with an opposite sign for the term
$v_{\mu}\overline{\psi}\gamma^{\mu}\psi,$ which implies breaking of
C-symmetry\footnote{One takes as starting point the Dirac equation $\left(
i\gamma^{\mu}\partial_{\mu}-e\gamma^{\mu}A_{\mu}-v_{\mu}\gamma^{\mu}-m\right)
\psi=0,$ which for an anti-particle must be rewritten with opposite charge
sign: $\left(  i\gamma^{\mu}\partial_{\mu}+e\gamma^{\mu}A_{\mu}-v_{\mu}%
\gamma^{\mu}-m\right)  \psi_{c}=0,$ being $\psi_{c}$ the anti-particle spinor.
In the case the $C$-symmetry holds on, this\ exact equation might be also
obtained by applying the charge conjugation operator $C=i\gamma^{0}\gamma^{2}$
on the initial Dirac equation. Making it, one attains: $\left(  i\gamma^{\mu
}\partial_{\mu}+e\gamma^{\mu}A_{\mu}+v_{\mu}\gamma^{\mu}-m\right)  \psi
_{c}=0,$ where one notes the opposite sign of the term $v_{\mu}\gamma^{\mu}$.
This puts in evidence the C- breakdown. A similar procedure may be employed to
demonstrate the conservation of PT symmetry.}.

One should now enquire about the spin interpretation of these solutions.
Obviously, such solutions will not present the same spin projection as the
usual Dirac free-particle solutions. But in some particular cases, it is
possible to show that such solutions exhibit the same spin projection. For
instance, whenever the background and the momentum are aligned along the
z-axis, the spinors take the form:
\begin{equation}
u_{1}=N\left(
\begin{array}
[c]{c}%
1\\
0\\
\frac{\left(  p_{z}-v_{z}\right)  }{E+m_{e}-v^{0}}\\
0
\end{array}
\right)  \ ,\ \ u_{2}=N\left(
\begin{array}
[c]{c}%
0\\
1\\
0\\
\frac{-\left(  p_{z}-a_{z}\right)  }{E+m_{e}-v^{0}}%
\end{array}
\right)  \ ,\text{ }v_{1}=N\left(
\begin{array}
[c]{c}%
\frac{\left(  p_{z}+v_{z}\right)  }{E+m_{e}+v^{0}}\\
0\\
1\\
0
\end{array}
\right)  ,\ \ v_{2}=N\left(
\begin{array}
[c]{c}%
0\\
\frac{-\left(  p_{z}+v_{z}\right)  }{E+m_{e}+v^{0}}\\
0\\
1
\end{array}
\right)  .
\end{equation}

Such solutions are eigenstates of the helicity operator, $\overrightarrow
{S}\cdot\widehat{p}=S_{z}=\frac{1}{2}\Sigma_{z}$, with: $\Sigma_{z}=\left(
\begin{array}
[c]{cc}%
\sigma_{z} & 0\\
0 & \sigma_{z}%
\end{array}
\right)  .$ Thus, the spinors $u_{1}$ and $v_{1}$ have eigenvalue $+1$ (spin
up) whereas the spinors $u_{2}$ and $v_{2}$ have eigenvalue $-1$ (spin down).
Hence, the presence of the fixed background does not suffice in principle to
change the spin polarization of the new states. A detailed study of the spin
projections may only be obtained by constructing the spin projector operators.
This point is addressed by Lehnert in ref. \cite{Hamilton}.

\subsection{Nonrelativistic limit}

Every good relativistic theory must exhibit a sensible low-energy limit whose
predictions may be compared with the results of other correlated
nonrelativistic theories. Such a requirement sets up the correspondence
between an intrinsically relativistic theory and a nonrelativistic one. In a
well-known case, the nonrelativistic limit of the Dirac theory yields the
Pauli equation, which consists of the Schr\"{o}dinger equation supplemented
with the spin-magnetic interaction. Hence, to work in the nonrelativistic
limit allows to investigate quantum mechanical features of a system without
losing relativistic effects (like spin) of the original theory. In the present
case, where the Dirac theory is being corrected by a Lorentz-violating
coupling term, one expects that the nonrelativistic regime be well described
by the Pauli equation incorporating Lorentz-violating terms. It will be shown
that this is exactly the case.

To correctly analyze the nonrelativistic limit of Lagrangian (\ref{L1}), this
model is considered in the presence of an external electromagnetic field
$\left(  A_{\mu}\right)  $, so that Lagrangian (\ref{L1}) is rewritten in the
form:
\begin{equation}
\mathcal{L=}\frac{1}{2}i\overline{\psi}\gamma^{\mu}\overleftrightarrow{D}%
_{\mu}\psi-m_{e}\overline{\psi}\psi-v_{\mu}\overline{\psi}\gamma^{\mu}\psi,
\label{DiracS3}%
\end{equation}
where $D_{\mu}=\partial_{\mu}+ieA_{\mu}.$ The external field is implemented
into our previous equations by means of the direct substitution: $p^{\mu
}\rightarrow p^{\mu}-eA^{\mu}.$ Replacing it into eqs. (\ref{WA}) and
(\ref{WB}), there follows: \qquad%
\begin{align}
w_{A}  &  =\frac{1}{\left(  E-eA^{0}-m_{e}-v^{0}\right)  }\overrightarrow
{\sigma}\cdot(\overrightarrow{p}-e\overrightarrow{A}-\overrightarrow{v}%
)w_{B},\label{WA2}\\
w_{B}  &  =\frac{1}{\left(  E-eA^{0}+m_{e}+v^{0}\right)  }\overrightarrow
{\sigma}\cdot(\overrightarrow{p}-e\overrightarrow{A}-\overrightarrow{v})w_{A}.
\label{WB2}%
\end{align}
$\ \ \ \ $\ \ \ 

In the low-velocity limit, it obviously holds $\left(  \overrightarrow
{p}\right)  ^{2}\ll m_{e}^{2},$ $eA_{0}\ll m_{e},$ conditions that impose the
smallness of kinetic and potential energy before the relativistic rest energy
$\left(  m_{e}\right)  $. With it, the energy of the system is written as
$E=m_{e}+H,$ where $H$ represents the nonrelativistic Hamiltonian. From eqs.
(\ref{WA2}) and (\ref{WB2}), the spinors $w_{A},w_{B}$ are read as the large
and the small components, once the magnitude of $w_{A}$ is much larger than
$w_{B}$. By replacing eq. $\left(  \text{\ref{WB2}}\right)  $ into eq.
(\ref{WA2}) and implementing the low-energy conditions, one should retain only
the equation for the strong component $\left(  w_{A}\right)  $,%

\begin{equation}
\left(  H-eA^{0}-v^{0}\right)  w_{A}=\frac{1}{\left(  2m_{e}+v^{0}\right)
}\overrightarrow{\sigma}\cdot(\overrightarrow{p}-e\overrightarrow
{A}-\overrightarrow{v})\overrightarrow{\sigma}\cdot(\overrightarrow
{p}-e\overrightarrow{A}-\overrightarrow{v})w_{A}, \label{HV1}%
\end{equation}
which describes the physics of the nonrelativistic limit. Using the identity,
$(\overrightarrow{\sigma}\cdot\overrightarrow{a})(\overrightarrow{\sigma}%
\cdot\overrightarrow{b})=\overrightarrow{a}\cdot\overrightarrow{b}%
+i\overrightarrow{\sigma}\cdot(\overrightarrow{a}\times\overrightarrow{b}),$
eq. (\ref{HV1}) is reduced to the form,
\begin{equation}
Hw_{A}=\biggl\{\frac{(\overrightarrow{p}-e\overrightarrow{A}-\overrightarrow
{v})^{2}}{2m_{e}}+\frac{1}{2m_{e}}\overrightarrow{\sigma}\cdot\lbrack
\nabla\times(\overrightarrow{A}-\overrightarrow{v})]+(eA^{0}+v^{0}%
)\biggr\}w_{A}, \label{HS}%
\end{equation}
where $H$ is the nonrelativistic Hamiltonian. Specifically, concerning the
spin-orbit interaction, one can see that such background does not yield any
modification, once $\nabla\times\overrightarrow{v}=0.$ Now, comparing Eq.
(\ref{HS}) with the Pauli equation, the Hamiltonian takes a more familiar
form:
\begin{equation}
H=\biggl\{\left[  \frac{(\overrightarrow{p}-e\overrightarrow{A})^{2}}{2m_{e}%
}-\frac{e\hbar}{2m}\overrightarrow{\sigma}\cdot\overrightarrow{B}%
+eA^{0}\right]  +\left[  -\frac{2(\overrightarrow{p}-e\overrightarrow{A}%
)\cdot\overrightarrow{v}}{2m_{e}}+v^{0}+\frac{\overrightarrow{v}^{2}}{2m_{e}%
}\right]  \biggr\}. \label{NONR}%
\end{equation}
The first term into brackets contains the well-known Pauli Hamiltonian,
whereas the second one is the correction Hamiltonian arising from the
Lorentz-violating background. This specific term, object of our attention, is
rewritten below:%

\begin{equation}
H_{LV}=\frac{2i\overrightarrow{v}\cdot\overrightarrow{\nabla}}{2m_{e}}%
+\frac{2e\overrightarrow{A}\cdot\overrightarrow{v}}{2m_{e}}+v_{0}%
+\frac{\overrightarrow{v}^{2}}{2m_{e}}. \label{HLV}%
\end{equation}

Here, note that the breakdown of charge conjugation is no more manifest, once
the relativistic dispersion relation has degenerated in a single expression
for particles and anti-particles. Looking at eq. (\ref{HLV}), the last two
terms change the nonrelativistic Hamiltonian only by a constant, which does
not represent any physical change (it just shifts the levels as a whole, not
modifying the transition energies). Thus, just the first and the second are
able to induce modifications on a physical system. The\textbf{ }purpose now is
to investigate the contribution of these two terms on the 1-particle wave
functions $\left(  \Psi\right)  $ of the hydrogen. It should be taken into
account only the first term, once the hydrogen atom is initially regarded as a
free system ($\overrightarrow{A}=0)$. \ This contribution is expected to be
null, once it represents an average of the linear momentum on an atomic bound
state. Explicitly, this energy quantity is correctly worked out as a first
order perturbation on the corresponding 1-particle wave functions, namely:
$\Delta E=\frac{i}{m_{e}}\langle nlm|\overrightarrow{v}{}\cdot\overrightarrow
{\nabla}|nlm\rangle,$ where $n,l,m$ are the usual quantum numbers that label
the 1-particle wave function for the hydrogen atom, $\Psi_{nlm}(r,\theta
,\phi)=R_{nl}(\rho)\Theta_{lm}(\theta)\Phi_{m}(\phi).$ Replacing such a form
in $\Delta E$, with the gradient operator written in spherical coordinates, it implies%

\begin{align}
\Delta E  &  =\frac{i}{m_{e}}\int\biggl\{R_{nl}\left(  r\right)  ^{\ast}%
\frac{\partial R_{nl}\left(  r\right)  }{\partial r}\left\vert \Theta
_{lm}\left(  \theta\right)  \right\vert ^{2}\left\vert \Phi_{m}\left(
\phi\right)  \right\vert ^{2}\overrightarrow{v}\cdot\widehat{r}+\frac
{\left\vert R_{nl}\left(  r\right)  \right\vert ^{2}\left\vert \Phi_{m}\left(
\phi\right)  \right\vert ^{2}}{r}\Theta_{lm}\left(  \theta\right)  ^{\ast
}\frac{\partial\Theta_{lm}\left(  \theta\right)  }{\partial\theta
}\overrightarrow{v}\cdot\widehat{\theta}\nonumber\\
&  +im\frac{\left\vert R_{nl}\left(  r\right)  \right\vert ^{2}\left\vert
\Theta_{lm}\left(  \theta\right)  \right\vert ^{2}\left\vert \Phi_{m}\left(
\phi\right)  \right\vert ^{2}}{r\sin\theta}\overrightarrow{v}\cdot
\widehat{\phi}\biggr\}d^{3}r.
\end{align}
For explicit calculation, the vector $\overrightarrow{v}$ can be placed along
the z-axis, so that: $\overrightarrow{v}\cdot\widehat{r}=v_{z}\cos
\theta,\overrightarrow{v}\cdot\widehat{\theta}=-v_{z}\sin\theta
,\overrightarrow{v}\cdot\widehat{\phi}=0.$ Thus, one notes that the first two
terms exhibit the presence of angular additional factors, $\cos\theta\ $and
$\sin\theta$, respectively. The first term is explicitly written as:%

\begin{equation}
\Delta E_{1}=\frac{iv_{z}}{m_{e}}\int\left[  R_{nl}\left(  r\right)  ^{\ast
}\frac{\partial R_{nl}\left(  r\right)  }{\partial r}\left\vert \Theta
_{lm}\left(  \theta\right)  \right\vert ^{2}\cos\theta\right]  r^{2}\sin\theta
drd\theta=0.
\end{equation}
This null result is a consequence of $\int_{0}^{\pi}\left[  \left\vert
\Theta_{lm}\left(  \theta\right)  \right\vert ^{2}\cos\theta\right]
\sin\theta d\theta=0,$ which holds for the associated Legendre functions.
Following, the second term
\begin{equation}
\Delta E_{2}=-\frac{iv_{z}}{m_{e}}\int\left[  \frac{\left\vert R_{nl}\left(
r\right)  \right\vert ^{2}}{r}\Theta_{lm}\left(  \theta\right)  ^{\ast}%
\frac{\partial\Theta_{lm}\left(  \theta\right)  }{\partial\theta}\sin
\theta\right]  r^{2}\sin\theta drd\theta,
\end{equation}
is now analyzed. The involved\textbf{ }angular integration reads as $\int
_{0}^{\pi}\Theta_{lm}\left(  \theta\right)  (\partial\Theta_{lm}%
/\partial\theta)\sin^{2}\theta d\theta=\int_{-1}^{1}[\Theta_{lm}\left(
z\right)  \partial\Theta_{lm}/\partial z](z^{2}-1)dz=0,$ which comes out null
as consequence of the recurrence relation, $(z^{2}-1)\frac{d}{dz}\Theta
_{lm}\left(  z\right)  =lz\Theta_{lm}\left(  z\right)  -(l+m)\Theta
_{l-1,m}\left(  z\right)  ,$ and of the following orthogonality relation:
$\int_{0}^{\pi}\Theta_{lm}\left(  z\right)  \Theta_{pm}(z)dz=0,$ for $l\neq
p.$ Therefore, the total energy correction is null, that is: $\Delta E=0.$
This means that the presence of the Lorentz-violating background does not
imply any energy shift in the hydrogen spectrum. It is instructive to claim
that this null correction is in full accordance with the role played by the
term $v_{\mu}\overline{\psi}\gamma^{\mu}\psi$: it only brings about a
4-momentum shift, $p^{\mu}\rightarrow p^{\mu}-v^{\mu},$ without any physical
consequence on the spectrum of the system.\textbf{ }It is also possible to
understand it by reading the effect of the background as a gauge
transformation. Indeed, making use of a field redefinition, $\psi
\rightarrow\Psi(x)=\psi(x)e^{-iv\cdot x},\ \ $it is possible to remove the
background from the theory, so that Lagrangian (\ref{L1}) takes on the usual
free form (written in terms of the field $\Psi$), namely: $\mathcal{L}%
\acute{}%
=\mathcal{L}_{Dirac}$. This is true in any theory containing only one fermion
field. For this result to remain valid in the case of a multifermion theory,
the fermions families should be uncoupled with each other (no interacting
fermions) and be coupled to the same Lorentz-violating parameter ($v_{\mu})$
\cite{Colladay1}-\cite{Colladay2}.

The general result provided by the relativistic spectrum of the hydrogen may
be attained by the exact solution of the modified Dirac equation
(\ref{DiracS1}), taken in the presence of the Coulombian potential. This
solution, however, will yield nothing new, once it corresponds exactly to the
conventional relativistic solution shifted according to $p^{\mu}\rightarrow
p^{\mu}-v^{\mu}$. Finally, it should be noted that this null outcome is not
due to the specific choice of the background spatial orientation, $v^{\mu
}=(v_{0},0,0,v_{z});$ by adopting a background along an arbitrary direction,
$\overrightarrow{v}=(v_{x},v_{y},v_{z}),$ identical calculations
straightforwardly yield the same null result for $\Delta E$.

So far, the hydrogen spectrum has been investigated only in the absence of
external field. In the presence of a fixed magnetic field, one notes that the
term $e\overrightarrow{A}\cdot\overrightarrow{v}/m_{e}$ of eq. (\ref{HLV}) may
contribute to a first order calculation by the quantity:
\begin{equation}
\Delta E_{A\cdot v}=\frac{e}{m_{e}}\int\Psi^{\ast}\left(  \overrightarrow
{A}\cdot\overrightarrow{v}\right)  \Psi d^{3}r.
\end{equation}
Knowing that $\overrightarrow{A}=-\overrightarrow{r}\times\overrightarrow
{B}/2,$ for a fixed magnetic field along the z-axis, $\overrightarrow{B}%
=B_{0}\widehat{z}$, it results: $\overrightarrow{A}=-B_{0}(y/2,-x/2,0).$ This
implies $\Delta E_{A\cdot v}=-(eB_{0}/2m_{e})\int\Psi^{\ast}\left(
yv_{x}-xv_{y}\right)  \Psi d^{3}r,$ whose explicit calculation leads to
$\Delta E_{A\cdot v}=0.$ Therefore, one concludes that the presence of a fixed
external magnetic field does not yield any Lorentz-violating contribution to
hydrogen spectrum besides the usual Zeeman effect.

It is instructive to remark that these calculations hold equivalently for the
case of a positron, for which the modified Pauli equation stems from $\left(
i\gamma^{\mu}\partial_{\mu}+e\gamma^{\mu}A_{\mu}+v_{\mu}\gamma^{\mu}-m\right)
\psi_{c}=0.$ In comparison with eq. (\ref{NONR}), the positron nonrelativistic
Hamiltonian exhibits opposite charge and opposite $v^{\mu}$ parameter,
implying a Lorentz-violating Hamiltonian in the form $H_{LV}%
=[-i\overrightarrow{v}\cdot\overrightarrow{\nabla}/m_{e}+e(\overrightarrow
{A}\cdot\overrightarrow{v})/m_{e}-v_{0}+\overrightarrow{v}^{2}/2m_{e}%
].$\textbf{ }However, as in the electron case, this Hamiltonian yields no
physically detectable energy shift. This issue is obviously related to the
analysis of the hydrogen and antihydrogen spectroscopy, realized in wide sense
in ref. \cite{Hydrog}. In this work, it is also taken into account the effect
of the Lorentz-violating background on the hyperfine structure (considering
the proton spin).

\section{Lorentz-violating Dirac Lagrangian ("axial vector" coupling)}

Amongst the possible coefficients involved with the breaking of Lorentz
symmetry in the fermion sector of the SME, shown in eq. (\ref{LV1}), our
interest rest in one that is also CPT-odd, $b_{\mu}\overline{\psi}\gamma
_{5}\gamma^{\mu}\psi.$ This torsion-like term \cite{Shapiro1}-\cite{Shapiro2}
is linked with the fixed background by means of an assigned axial vector
coupling. Taking it into account, one writes:%

\begin{equation}
\mathcal{L}=\frac{1}{2}i\overline{\psi}\gamma^{\mu}\overleftrightarrow
{\partial}_{\mu}\psi-m_{e}\overline{\psi}\psi-b_{\mu}\overline{\psi}\gamma
_{5}\gamma^{\mu}\psi. \label{LPS}%
\end{equation}
The first step is to determine the new Dirac equation stemming from the above
Lagrangian, namely:%

\begin{equation}
\left(  i\gamma^{\mu}\partial_{\mu}-b_{\mu}\gamma_{5}\gamma^{\mu}%
-m_{e}\right)  \psi=0. \label{Dirac2}%
\end{equation}
This modified equation is then rewritten in the momentum space,%

\begin{equation}
\left(  \gamma^{\mu}p_{\mu}-b_{\mu}\gamma_{5}\gamma^{\mu}-m_{e}\right)
w(p)=0, \label{Dirac3}%
\end{equation}
provided a plane-wave solution is proposed. In order to obtain the dispersion
relation associated with such an equation, it should be multiplied by $\left(
\gamma^{\mu}p_{\mu}-b_{\mu}\gamma_{5}\gamma^{\mu}+m_{e}\right)  ,$ so that one
obtains: $[p^{2}-m_{e}^{2}-b^{2}+\gamma_{5}%
({\rlap{\hbox{$\mskip 1 mu /$}}p\rlap{\hbox{$\mskip 1 mu /$}}b-\rlap{\hbox{$\mskip 1 mu /$}}b\rlap{\hbox{$\mskip 1 mu /$}}p)]w(p)=0.}%
$ This expression presents contributions out of the main diagonal of the
spinor space. In order to achieve an expression totally contained in the main
diagonal, equally valid for each component of the spinor ${w,}$ the preceding
equation shall be multiplied by $(p^{2}-m_{e}^{2}-b^{2}-\gamma_{5}%
({\rlap{\hbox{$\mskip 1 mu /$}}p\rlap{\hbox{$\mskip 1 mu /$}}b-\rlap{\hbox{$\mskip 1 mu /$}}b\rlap{\hbox{$\mskip 1 mu /$}}p),}%
$ which yields the following dispersion relation:
\[
(p^{2}-m_{e}^{2}-b^{2})^{2}+4p^{2}b^{2}-4(p\cdot b)^{2}=0.
\]
This is a fourth order relation for the energy that can be exactly solved only
in special cases. In the case of a purely timelike background, $b^{\mu}%
=(b_{0},0),$ and a purely spacelike background, $b^{\mu}=(0,\overrightarrow
{b}),$ one respectively achieves:
\begin{align}
E  &  =\pm\sqrt{\overrightarrow{p}^{2}+m_{e}^{2}+b_{0}^{2}\pm2b_{0}%
|\overrightarrow{p}|},\label{Energy3}\\
E  &  =\pm\sqrt{\overrightarrow{p}^{2}+m_{e}^{2}+\overrightarrow{b}^{2}%
\pm2\left[  m_{e}^{2}\overrightarrow{b}^{2}+(\overrightarrow{b}\cdot
\overrightarrow{p})^{2}\right]  ^{1/2}}. \label{Energy5}%
\end{align}
Notice that there is no breakdown of charge conjugation in this case. In fact,
after usual reinterpretation both particle and anti-particle exhibit the same
energy values, that is, the positive roots given in eqs. (\ref{Energy3}),
(\ref{Energy5}). Therefore, Lagrangian (\ref{LPS}) does not imply C-violation.
This may be explicitly demonstrated by means of the procedure employed in
Footnote 2.

Taking into account the $\gamma-$matrices definition, given at footnote 1,
eq$.\left(  \text{\ref{Dirac2}}\right)  $ gives rise to two coupled spinor
equations for $w_{A}$ and $w_{B}$:
\begin{align}
\left(  E-\overrightarrow{\sigma}{}\cdot\overrightarrow{b}-m_{e}\right)
w_{A}+\left(  b^{0}-\overrightarrow{\sigma}{}\cdot\overrightarrow{p}\right)
w_{B}  &  =0,\label{SP1}\\
\left(  \overrightarrow{\sigma}{}\cdot\overrightarrow{p}-b^{0}\right)
w_{A}+\left(  -E+\overrightarrow{\sigma}{}\cdot\overrightarrow{b}%
-m_{e}\right)  w_{B}  &  =0, \label{SP2}%
\end{align}
\ \ leading to the following spinor relations:%

\begin{align}
w_{A}  &  =\frac{1}{E_{2}^{2}}\biggl\{\left(  E-m_{e}\right)  \left(
\overrightarrow{\sigma}\cdot\overrightarrow{p}\right)  -\left(  E-m_{e}%
\right)  b^{0}-b^{0}(\overrightarrow{\sigma}\cdot\overrightarrow
{b})+\overrightarrow{b}\cdot\overrightarrow{p}+i\overrightarrow{\sigma}%
\cdot\overrightarrow{c}\biggr\}w_{B},\label{WAT}\\
w_{B}  &  =\frac{1}{E_{1}^{2}}\biggl\{\left(  E+m_{e}\right)  \left(
\overrightarrow{\sigma}\cdot\overrightarrow{p}\right)  -\left(  E+m_{e}%
\right)  b^{0}-b^{0}(\overrightarrow{\sigma}\cdot\overrightarrow
{b})+\overrightarrow{b}\cdot\overrightarrow{p}+i\overrightarrow{\sigma}%
\cdot\overrightarrow{c}\biggr\}w_{A}, \label{WBT}%
\end{align}
where: $\overrightarrow{c}=\overrightarrow{b}\times\overrightarrow{p}%
,E_{1}^{2}=\left[  \left(  E+m_{e})^{2}-b\cdot b\right)  \right]  ,E_{2}%
^{2}=\left[  \left(  E-m_{e})^{2}-b\cdot b\right)  \right]  .$

To construct the plane-wave solutions, one follows the general procedure
adopted in the preceding section. The resulting $4\times1$ spinor solutions
are given below:%

\begin{align}
u_{1}  &  =N\left(
\begin{array}
[c]{c}%
1\\
0\\
\left[  \left(  E+m_{e}\right)  \left(  p_{z}-b^{0}\right)  -b^{0}%
b_{z}+\overrightarrow{b}\cdot\overrightarrow{p}+ic_{z}\right]  /E_{1}^{2}\\
\left[  \left(  E+m_{e}\right)  \left(  p_{x}+ip_{y}\right)  -b^{0}\left(
b_{x}+ib_{y}\right)  +i\left(  c_{x}+ic_{y}\right)  \right]  /E_{1}^{2}%
\end{array}
\right)  \ ,\ \ \\
\ u_{2}  &  =N\left(
\begin{array}
[c]{c}%
0\\
1\\
\left[  \left(  E+m_{e}\right)  \left(  p_{x}-ip_{y}\right)  -b^{0}\left(
b_{x}-ib_{y}\right)  +i\left(  c_{x}-ic_{y}\right)  \right]  /E_{1}^{2}\\
\left[  -\left(  E+m_{e}\right)  \left(  p_{z}+b^{0}\right)  +b^{0}%
b_{z}+\overrightarrow{b}\cdot\overrightarrow{p}-ic_{z}\right]  /E_{1}^{2}%
\end{array}
\right)  ,
\end{align}

\begin{align}
v_{1}  &  =N\left(
\begin{array}
[c]{c}%
\left[  \left(  E+m_{e}\right)  (p_{z}+b^{0})+b^{0}b_{z}+\overrightarrow
{b}\cdot\overrightarrow{p}-ic_{z}\right]  /E_{2}^{2}\\
\left[  \left(  E+m_{e}\right)  \left(  p_{x}+ip_{y}\right)  -b^{0}\left(
b_{x}+ib_{y}\right)  -i\left(  c_{y}+ic_{z}\right)  \right]  /E_{2}^{2}\\
1\\
0
\end{array}
\right)  \ ,\ \ \ \ \\
v_{2}  &  =N\left(
\begin{array}
[c]{c}%
\left[  \left(  E+m_{e}\right)  \left(  p_{x}-ip_{y}\right)  +b^{0}\left(
b_{x}-ib_{y}\right)  -i\left(  c_{x}-ic_{y}\right)  \right]  /E_{2}^{2}\\
\left[  -\left(  E+m_{e}\right)  \left(  p_{z}-b^{0}\right)  -b^{0}%
b_{z}+\overrightarrow{b}\cdot\overrightarrow{p}+ic_{z}\right]  /E_{2}^{2}\\
0\\
1
\end{array}
\right)  ,
\end{align}
where N is the normalization constant. The eigenvalues of energy are the ones
evaluated in eqs. (\ref{Energy3}), (\ref{Energy5}) that are now exhibited in
the following eigenenergy relations: $Hu_{i}=E_{i}^{(u)}u_{i},$ with
$E_{i}^{(u)}=\left[  \overrightarrow{p}^{2}+m_{e}^{2}+b_{0}^{2}+(-1)^{i}%
2b_{0}|\overrightarrow{p}|\right]  ^{1/2}$, for $b^{\mu}=(b_{0},0),$ and
$E_{i}^{(v)}=\left[  \overrightarrow{p}^{2}+m_{e}^{2}+\overrightarrow{b}%
^{2}+(-1)^{i}2[m_{e}^{2}\overrightarrow{b}^{2}+(\overrightarrow{b}%
\cdot\overrightarrow{p})^{2}]^{1/2}\right]  ^{1/2},$ for $b^{\mu
}=(0,\overrightarrow{b}),$ and $i=1,2.$ Here, $E_{i}^{(u)}$ stands for the
particle and anti-particle energy. Despite the cumbersome form of these
spinors, it is possible to show that in the case of $b^{\mu}=(b_{0}%
,0,0,b_{z})$ and $p=(0,0,p_{z}),$ such solutions are eigenstates of the spin
operator $\Sigma_{z}$\ with eigenvalues $\pm1,$ in much the same way as
observed in the foregoing section.

\subsection{Nonrelativistic limit}

The nonrelativistic limit of the model described by Lagrangian (\ref{LPS}) is
now worked out in much the same way of the previous section. The objective is
to identify the corrected Hamiltonian and possible energy shifts induced on
the spectrum of hydrogen in the presence and absence of an external magnetic
field. Considering the presence of an external electromagnetic field minimally
coupled to the spinor field:%

\begin{equation}
\mathcal{L=}\frac{1}{2}i\overline{\psi}\gamma^{\mu}\overleftrightarrow{D}%
_{\mu}\psi-m_{e}\overline{\psi}\psi-b_{\mu}\overline{\psi}\gamma_{5}%
\gamma^{\mu}\psi, \label{Ldirac}%
\end{equation}
where $D_{\mu}=\partial_{\mu}+ieA_{\mu}.$ Taking into account the external
field, eqs. (\ref{SP1}) and (\ref{SP2}) take on the form:%

\begin{align}
\left[  E-\overrightarrow{\sigma}{}\cdot\overrightarrow{b}-m_{e}%
-eA_{0}\right]  w_{A}+\left[  b^{0}-\overrightarrow{\sigma}{}\cdot
(\overrightarrow{p}-e\overrightarrow{A})\right]  w_{B}  &  =0,\label{WA4}\\
\left[  \overrightarrow{\sigma}{}\cdot(\overrightarrow{p}-e\overrightarrow
{A})-b^{0}\right]  w_{A}-\left[  E-\overrightarrow{\sigma}{}\cdot
\overrightarrow{b}+m_{e}-eA_{0}\right]  w_{B}  &  =0. \label{WB4}%
\end{align}
\ \ 

The low-energy limit is implemented by the following conditions: $\left(
\overrightarrow{p}\right)  ^{2}\ll m_{e}^{2},$ $eA_{0}\ll m_{e},$ $E=m_{e}+H.$
Furthermore, one still assumes that the factor $\overrightarrow{\sigma}%
\cdot\overrightarrow{b}$ must be neglected in eq. (\ref{WB4}), once the
background is supposed to be small whenever compared with the electron mass.
Implementing all these conditions, it holds for the strong component:
\qquad\qquad%

\begin{equation}
Hw_{A}=\biggl\{\left[  \overrightarrow{\sigma}{}\cdot(\overrightarrow
{p}-e\overrightarrow{A})\overrightarrow{\sigma}{}\cdot(\overrightarrow
{p}-e\overrightarrow{A})-2b_{0}\overrightarrow{\sigma}{}\cdot(\overrightarrow
{p}-e\overrightarrow{A})+b_{0}^{2}\right]  /2m_{e}+eA_{0}+\overrightarrow
{\sigma}{}\cdot\overrightarrow{b}\biggr\}w_{A}, \label{Pauli2}%
\end{equation}
After some algebraic calculations, one achieves:%

\begin{equation}
H=H_{\text{Pauli}}+\left[  \overrightarrow{\sigma}{}\cdot\overrightarrow
{b}-2b_{0}\overrightarrow{\sigma}{}\cdot(\overrightarrow{p}-e\overrightarrow
{A})/2m_{e}+b_{0}^{2}/2m_{e}\right]  . \label{HLV2}%
\end{equation}
This is the modified full Hamiltonian, composed by the Pauli and a
Lorentz-violating part $\left(  H_{LV}\right)  $, where in lies our interest.
Provided that $H_{LV}$ has two interesting new terms (the third one is
constant), one should try to figure out whether these terms imply real
corrections to the spectrum of hydrogen. Taking into account these
informations, the effective Lorentz-violating Hamiltonian assumes the form:
$H_{LV}=\overrightarrow{\sigma}{}\cdot\overrightarrow{b}-2b_{0}%
(\overrightarrow{\sigma}{}\cdot\overrightarrow{p})/2m_{e}$, where it was taken
$\overrightarrow{A}=0.$ One then starts analyzing the term $\overrightarrow
{\sigma}{}\cdot\overrightarrow{b},$ whose first order contribution is:%

\begin{equation}
\Delta E_{\sigma\cdot b}=\langle nljm_{j}m_{s}|\overrightarrow{\sigma}{}%
\cdot\overrightarrow{b}|nljm_{j}m_{s}\rangle. \label{EB}%
\end{equation}
Here, $n,l,j,m_{j}$ are the quantum numbers suitable to address a situation
where occurs addition of angular momenta ($L$ and $S$). To solve this
calculation, it is necessary to write the $|jm_{j}\rangle$ kets in terms of
the spin eigenstates $|mm_{s}\rangle,$ which is done by means of the general
expression: $|jm_{j}\rangle=%
{\displaystyle\sum\limits_{m,m_{s}}}
\langle mm_{s}|jm_{j}\rangle$ $|mm_{s}\rangle,$ where $\langle mm_{s}%
|jm_{j}\rangle$ are the Clebsch-Gordon coefficients. Evaluating such
coefficients for the case $j=l+1/2,m_{j}=m+1/2,$ one has: $\ |jm_{j}%
\rangle=\alpha_{1}|m\uparrow\rangle+\alpha_{2}|m+1\downarrow\rangle;$ one the
other hand, for $j=l-1/2,m_{j}=m+1/2,$ it results: $|jm_{j}\rangle=\alpha
_{2}|m\uparrow\rangle-\alpha_{1}|m+1\downarrow\rangle,$ with: $\alpha
_{1}=\sqrt{(l+m+1)/(2l+1)},\alpha_{2}=\sqrt{(l-m)/(2l+1)}.$ Now, taking into
account the orthonormalization relation $\langle m^{\prime}m_{s}^{\prime
}|mm_{s}\rangle=\delta_{m^{\prime}m}\delta_{m_{s}^{\prime}m_{s}},$ it is
possible to show that eq.\ (\ref{EB}) reduces simply to $\Delta E_{\sigma\cdot
b}=\langle jm_{j}|\sigma_{z}{}b_{z}|jm_{j}\rangle,$ whose explicit calculation
leads to:%

\begin{equation}
\Delta E_{\sigma\cdot b}=\pm\frac{b_{z}m_{j}}{2l+1}, \label{Zeeman}%
\end{equation}
where the positive and negative signs correspond to $j=l+1/2$ and $j=l-1/2,$
respectively. Thus, in this first order evaluation the energy turns out
corrected by a quantity depending on $\pm m_{j}$, in a very similar way to the
well-known Zeeman effect. Indeed, each line of the spectrum is split into
($2j+1$) lines, with a $b_{z}/(2l+1)$ linear separation. This correction was
also obtained in ref. \cite{Shore}. Once the magnitude of such splitting
depends directly on the modulus of the background, this theoretical outcome
may be used to set up an upper bound on the breaking parameter $\left(
b^{\mu}\right)  $.

Next, one evaluates the first order contribution of the second term of
$H_{LV}$ to the hydrogen spectrum, namely:%
\begin{equation}
\Delta E_{\sigma\cdot p}=\frac{ib_{0}}{m_{e}}\langle nljm_{j}m_{s}%
|\overrightarrow{\sigma}{}\cdot\overrightarrow{\nabla}|nljm_{j}m_{s}\rangle,
\label{EP}%
\end{equation}
The 1-particle wave function,$\Psi_{nljm_{j}m_{s}}=\psi_{nljm_{j}}%
(r,\theta,\phi)\chi_{sm_{s}},$ now contains a spin function, $\chi_{sm_{s}}.$
In order to solve eq. (\ref{EP}), \ one should note that the gradient operator
acts on the spatial function $\psi_{nljm_{j}}$, whereas $\overrightarrow
{\sigma}{}$ operates on the spin function, so that it reads:%

\begin{align}
&  =\frac{ib_{0}}{m_{e}}\int\biggl\{R_{nl}\left(  r\right)  ^{\ast}%
\frac{\partial R_{nl}\left(  r\right)  }{\partial r}\left\vert \Theta
_{lm}\left(  \theta\right)  \right\vert ^{2}\left\vert \Phi_{m}\left(
\phi\right)  \right\vert ^{2}\langle jm_{j}|\overrightarrow{\sigma}%
\cdot\widehat{r}|jm_{j}\rangle+\frac{\left\vert R_{nl}\left(  r\right)
\right\vert ^{2}\left\vert \Phi_{m}\left(  \phi\right)  \right\vert ^{2}}%
{r}\Theta_{lm}\left(  \theta\right)  ^{\ast}\nonumber\\
&  \frac{\partial\Theta_{lm}\left(  \theta\right)  }{\partial\theta}\langle
jm_{j}|\overrightarrow{\sigma}\cdot\widehat{\theta}|jm_{j}\rangle
+\frac{\left\vert R_{nl}\left(  r\right)  \right\vert ^{2}\left\vert
\Theta_{lm}\left(  \theta\right)  \right\vert ^{2}\left\vert \Phi_{m}\left(
\phi\right)  \right\vert ^{2}}{r\sin\theta}\langle jm_{j}|(\overrightarrow
{\sigma}\cdot\widehat{\phi})|jm_{j}\rangle\biggr\}d^{3}r.
\end{align}

Writing the spherical versors in terms of the Cartesian ones, one
obtains:$\overrightarrow{\sigma}\cdot\widehat{r}=\sin\theta\cos\phi\sigma
_{x}+\sin\theta\sin\phi\sigma_{y}+\cos\theta\sigma_{z},$ $\overrightarrow
{\sigma}\cdot\widehat{\theta}=\cos\theta\cos\phi\sigma_{x}+\cos\theta\sin
\phi\sigma_{y}-\sin\theta\sigma_{z},$ $\overrightarrow{\sigma}\cdot
\widehat{\phi}=-\sin\phi\sigma_{x}+\cos\phi\sigma_{y}.$ It is clear that only
the terms proportional to $\sigma_{z}$ yield non-null expectation values on
the kets $|jm_{j}\rangle$, which implies:
\begin{equation}
\Delta E_{\sigma\cdot p}=\frac{\pm ib_{0}m_{j}}{(2l+1)m_{e}}\int
\biggl\{R_{nl}^{\ast}\left(  r\right)  \frac{\partial R_{nl}}{\partial
r}\left\vert \Theta_{lm}\left(  \theta\right)  \right\vert ^{2}\cos
\theta-\frac{\left\vert R_{nl}\right\vert ^{2}}{r}\Theta_{lm}^{\ast}\left(
\theta\right)  \frac{\partial\Theta_{lm}}{\partial\theta}\sin\theta
\biggr\}d^{3}r.
\end{equation}
These are exactly the same integrals involved in the expressions of $\Delta
E_{1}$ and $\Delta E_{2},$ already evaluated in the previous section. So, it
is obvious that: $\Delta E_{\sigma\cdot p}=0.$ Hence, the sole non-null first
order effect on the hydrogen spectrum is a Zeeman-like splitting stemming from
the correction term $\overrightarrow{\sigma}{}\cdot\overrightarrow{b}.$

Another point that deserves attention is related to the correction term
$2eb_{0}\overrightarrow{\sigma}{}\cdot\overrightarrow{A},$ present in eq.
(\ref{HLV2}). This term is obviously null for the "free" hydrogen atom (once
$\overrightarrow{A}=0).$ For the case the atom is subjected to the influence
of an external magnetic field, however, this term must be taken into account.
For a fixed magnetic field along the z-axis, $\overrightarrow{B}=B_{0}%
\widehat{z}$, one has $\overrightarrow{A}=-B_{0}(y/2,-x/2,0),$ so that the
correction may be written as:
\begin{equation}
\Delta E_{\sigma\cdot A}=\frac{b_{0}e}{m_{e}}\langle nljm_{j}m_{s}%
|\overrightarrow{\sigma}{}\cdot\overrightarrow{A}|nljm_{j}m_{s}\rangle
=-\frac{B_{0}b_{0}e}{2m_{e}}\langle nljm_{j}m_{s}|y\sigma_{x}-x\sigma
_{y}|nljm_{j}m_{s}\rangle.
\end{equation}
Considering the effect of the spin operators on the kets $|jm_{j}\rangle$, a
null correction $\left(  \Delta E_{\sigma\cdot A}=0\right)  $ turns out. One
should remark that this result remains null even for an arbitrary orientation
of the magnetic field. Therefore, the conclusion is that an external fixed
field does not imply any additional correction to the well-known Zeeman
effect. In this case, the Lorentz-violating effect of eq. (\ref{Zeeman})
corrects the usual Zeeman splitting just by a small quantity proportional to
$|\overrightarrow{b}|$. The general result provided by the relativistic
spectrum of the hydrogen may be examined by the exact solution of the modified
Dirac equation (\ref{DiracS1}) in the presence of the Coulombian potential.
This case implies qualitative modifications on the usual relativistic hydrogen
spectrum, both in the case of a purely timelike or purely spacelike
background. It is now under development.

\section{Conclusion}

In this work, the effects of CPT- and Lorentz-violating background terms
(stemming from a more fundamental theory) on the Dirac equation have been
studied. This analysis has considered two different ways of coupling the
fermion field to the background. One has started with the vector coupling, for
which the modified Dirac equation with corresponding solutions and
eigenenergies have been determined. The results agree with those already known
in the literature \cite{Colladay1}-\cite{Colladay2}, \cite{Hamilton}%
-\cite{Hamilton2}. The nonrelativistic regime has been assessed. It was
verified that the background implies modifications on the Pauli equation, but
they are such that do not yield any energy shift for the hydrogen spectrum.
This is an expected result, once the vector coupling might be seen simply as a
\ momentum shift ($p^{\mu}\rightarrow p^{\mu}-v^{\mu})$ unable to bring about
physical modifications, or as a gauge transformation that absorbs entirely the
background. In the sequel, one has analyzed the case in which the background
is coupled to spinor field in an axial vector way. Again, the free-particle,
dispersion relation and eigenenergies have been calculated and the
nonrelativistic limit has been discussed. It was argued that the
Lorentz-violating corrections to the Pauli equation are able to provide new
effects on the spectrum of hydrogen. Indeed, it has been shown that the
background may induce a Zeeman-like splitting of the spectral lines arising
from a spin interaction. This effect may be used to set up bounds on the
magnitude of the Lorentz-violation coefficient, $b^{\mu},$ according to
precise observations of hydrogen spectrum. The presence of an external
homogenous magnetic field has been also considered, but it has been shown that
it does not add new corrections on the usual Zeeman effect beyond the ones
already\ associated with the coefficient $b_{\mu}.$

Further comments refer to the possibility of inducing topological phases in
the electron wave function by the Lorentz-violating terms considered. In a
recent work \cite{Thales}, it has been argued that the fixed background,
whenever non-minimally coupled to the gauge and spinor fields by means of a
Carroll-Field-Jackiw-like term, $\epsilon_{\mu\nu\alpha\beta}\gamma^{\mu
}v^{\nu}F^{\alpha\beta},$ is able to induce an Aharonov-Casher phase in the
wave function of an electron. This occurs whenever the canonical momentum is
changed by a term whose curl is non null. In the case of the CPT- and
Lorentz-violating coupling terms investigated in this work, however, no
topological phase is generated. In effect, in both cases the canonical
momentum is changed by a constant quantity ($\overrightarrow{p}\rightarrow
\overrightarrow{p}-\overrightarrow{v}$\ ) or remains invariant. In this paper,
however, it was not addressed the possible effects\ induced by the
non-minimally coupled background on a low-energy atomic spectrum. This issue
has been just recently addressed \cite{Fernando1}, with new interesting
results. Another continuation of the\ present line of investigation consists
in examining the solution of the full Lorentz-violating relativistic Dirac
equation for an interacting configuration, such as the Coulombian potential.
In this case, only the axial vector coupling should be considered, once the
vector coupling just implies a momentum shift unable to modify the solutions.
One then expects that the relativistic spectrum solution may reveal new
effects, at the same time it recovers the results here evaluated in the
nonrelativistic regime.

\begin{acknowledgments}
The authors express their gratitude both to FAPEMA\ (Funda\c{c}\~{a}o de
Amparo \`{a} Pesquisa do Estado do Maranh\~{a}o) and to the CNPq (Conselho
Nacional de Desenvolvimento Cient\'{\i}fico e Tecnol\'{o}gico) for financial support.
\end{acknowledgments}

\end{document}